# Graphene p-n junction arrays as quantum-Hall resistance standards

M. Woszczyna, M. Friedemann, T. Dziomba, Th. Weimann, and F. J. Ahlers[a]
*Physikalisch-Technische Bundesanstalt, Bundesallee 100, D-38116 Braunschweig, Germany*

We demonstrate a device concept to fabricate resistance standards made of quantum Hall series arrays by using p-type and n-type graphene. The ambipolar nature of graphene allows fabricating series quantum Hall resistors without complex multi-layer metal interconnect technology, which is required when using conventional GaAs two-dimensional electron systems. As a prerequisite for a precise resistance standard we confirm the vanishing of longitudinal resistance across a p-n junction for metrological relevant current levels in the range of a few microamperes.

Graphene is an electronic material which, since its discovery in 2004, has triggered an avalanche of theoretical and experimental studies.[1] Its band structure gives rise to a number of fascinating properties, making it a promising material for next generation electronic devices.[2] Especially for electrical metrology graphene has unique advantages: since the quantum Hall effect persists up to room temperature,[3] and since the Landau level spacing in a small magnetic field $B$ decreases only with the square root of $B$, graphene offers the exciting possibility of a resistance standard working at 4.2 K or higher, and in a magnetic field of only 1 or 2 Tesla.[4,5] In metrology the quantum Hall effect is used to realize a value of electrical resistance with relative measurement uncertainty of a few parts in $10^9$ or better, typically employing GaAs based heterostructures hosting a 2-dimensional electron system (2DES). Parallel or series quantum Hall arrays could cover a wider resistance scale than just the singular value of 12.9 k$\Omega$. Such quantum Hall arrays are technically feasible, but they are not used in practice, mainly due to the technical difficulties to produce arrays which reliably allow a low measurement uncertainty. Here we present a device concept which avoids these difficulties by exploiting the unique feature of graphene that it can support a 2-dimensional hole system (2DHS) as well as a 2-dimensional electron system in the same device. We support our concept by demonstrating that the prerequisite for a quantum Hall series resistance standard, the vanishing of longitudinal resistance across the series connection, is met even at the high current levels required in practice.

The technique of connecting quantum Hall devices in series or in parallel, to obtain multiples or fractions of the resistance $h/2e^2$, was first demonstrated by Delahaye.[6] For the case of a series connection, the principle is illustrated in FIG. 1(a). Twice the value of the single-Hall-bar resistance $R_H = h/2e^2$ is measured between terminals 5 and 5' because the voltage drop in the connection (2-2') is practically zero as can be shown by an equivalent circuit model of a quantum Hall device.[7] The model predicts that in Hall-bars with multiple inter-connections the current in each additional connection is smaller than in the preceding one by a factor $\varepsilon/(\varepsilon+2)$ with $\varepsilon = R_c/R_H$, where $R_c$ is the interconnect resistance. When $\varepsilon \ll 1$, three interconnects between successive Hall bars already suffice to make the voltage drop $V_{2-2'}$ so small that $V_{5-5'} = V_{5-2} + V_{2'-5'} + (\varepsilon/2)^3 V_{2-2'} = 2 \cdot h/2e^2$, since $(\varepsilon/2)^3$ is practically negligible. With $M$ Hall bars in series a resistance $M \cdot h/2e^2$ is obtained.[8] Corresponding devices have been made, e.g. arrays comprising 10 Hall bars in series to produce fundamental constant based resistance standards of 129 k$\Omega$.[9] However, since metallic leads must cross each other, a multilayer interconnect technology is required which complicates fabrication. Failure of the insulation between crossing interconnects would cause an additional Hall voltage contribution from the lead metal layer, compromising the achievable uncertainty. Further, all contacts between metal leads and Hall bars must be of low resistance in order to ensure low $\varepsilon$-values. For these reasons series arrays made from GaAs based two-dimensional systems are not yet routinely used in standards laboratories. With graphene, however, Hall bars of p-type and n-type can be combined. This requires no crossing of interconnects, as is illustrated in FIG. 1(b), where the series connection scheme is shown for inverted carrier polarity in the second Hall bar. The possibility to avoid multi-layer technology would already justify the development of p-type and n-type graphene series arrays, but as we show below one can go even further and get rid of interconnects altogether by realizing the scheme of FIG. 1(c).

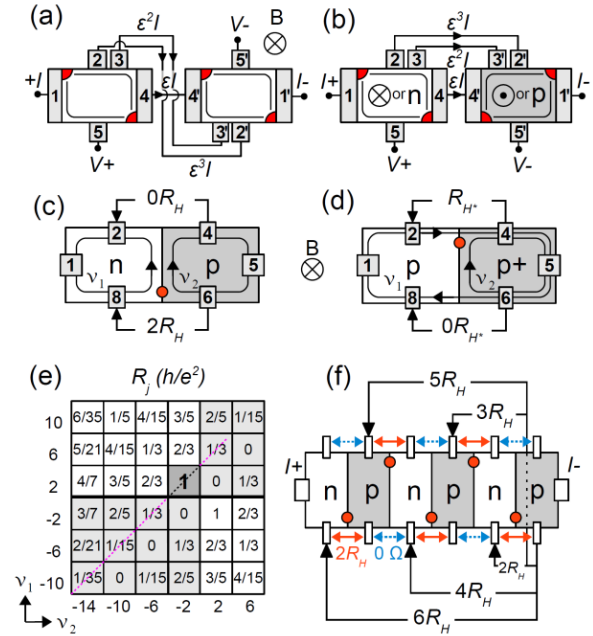

FIG. 1. (Color online) Triple series connection scheme of two Hall bars in standard configuration (a), and for the case of opposite magnetic field or carrier type (b). Note that in (b) crossings of the wire leads can be avoided. Areas of high current density ('hot spots') are marked in device corners; lines within the Hall bars denote equi-potential lines. (c) Directly connected Hall bars with bipolar filling (c) and unipolar, but unequal filling (d). In (c) and (d) lines denote edge states, not equi-potential lines. (e) Theoretical resistance values calculated for samples (c) or (d). (f) Schematic diagram of the graphene p-n junction series array.

---

[a] Author to whom correspondence should be addressed. Electronic mail: Franz.Ahlers@ptb.de.

In order to demonstrate this we fabricated a Hall bar consisting of two sub-bars of different carrier polarity. From natural graphite a graphene flake was exfoliated onto Si/SiO$_2$. The flake was shaped into a Hall bar geometry by Ar-ion etching, and Ti/Au contacts (10 nm/50 nm) were evaporated. One half of the device was covered with HfO$_2$ by thermal evaporation from a white HfO$_2$ tablet under $8\times10^{-5}$ mbar oxygen pressure. Atomic force microscopy imaging revealed homogeneous coverage of the graphene device, see image in FIG. 2(a). An *RMS* roughness of the HfO$_2$ edge of about 23 nm and an HfO$_2$ layer thickness of 125 nm were obtained. The room temperature resistance dependence on back gate voltage shown in Fig. 2(c) confirms that the Hall bar indeed comprises two sub-bars of different carrier concentration. The two-terminal resistance between contacts 1 and 5 results from the series combination of the sub-bars whose individual four-terminal resistances in the two lower curves were measured between contact pairs 2-3 (left curve) and 3-4 (right curve). The sample was kept in vacuum of approximately $10^{-6}$ mbar for two hours at room temperature before carrying out the measurements, and a DC measurement current of 1 μA was used. The HfO$_2$ layer caused additional p-type doping of the graphene, shifting the graphene charge neutrality point (CNP) towards higher voltages. The CNP voltage difference between uncovered and covered graphene regions was 15.8 V, corresponding to an additional acceptor doping of about $1.13\times10^{12}$ cm$^{-2}$. At low temperature (50 mK) the CNP voltage in the uncovered part decreased by 7 V and the CNP voltage of the covered part increased by 1.6 V. This effect may be due to continuous cleaning of the open graphene surface during measuring, while contaminations trapped under HfO$_2$ could not be removed. Neglecting electron-hole puddle phenomena around the CNPs,[10] two carrier types take part in transport within the back gate voltage range between 45.5 V to 69.2 V, whereas for lower or higher voltages only one type exists.

That a series p-n junction does indeed act like the double device of FIG. 1(b), but with the discrete and finite number of interconnects replaced by the continuous interconnection of FIG. 1(c), becomes evident from the measurements in FIG 3. It presents four-terminal resistance measurements in dependence on back-gate voltage at magnetic field $B$ = 18 T and at temperature $T$ = 50 mK. Hall resistances across the uncovered and covered graphene parts and resistances along both sides of the Hall bar were recorded simultaneously in one voltage scan. The connection scheme is drawn in the inset. Unlike in all previous investigations of p-n junctions, where only ac-currents in the nA range were applied,[11-13] we used high dc-current levels of up to 9 μA, as are required in precision resistance calibrations. In the individual uncovered and covered parts we observed standard Hall resistance quantization sequences (black dotted levels show theoretical plateaus at filling factors ν = ±2 and ν = ±6). At voltages between 45.5 V and 69.8 V (shaded area in FIG 3(a)) the Hall bar consists of p-type and n-type sub-bars in series and the Hall voltages in the two parts have opposite sign. To the left and right of this interval carriers have the same polarity in both sub-bars. The resistances at contact pairs 8-6 and 2-4 exhibit Hall plateaus corresponding to resistances $h/e^2$ (i.e. twice the maximum value of $h/2e^2$ of a single bar), $h/3e^2$, and $h/15e^2$, as well as the zero resistance plateaus expected for such nominally 'longitudinal' measurement geometry.

The quantized Hall resistance values obtained in such measurements have been qualitatively explained in the framework of the Landauer-Büttiker edge-state transport model (ESM).[14,15] In the p-n junction case, the edge channels circulate along the boundary in opposite directions. They begin to equilibrate when they meet at the p-n interface and become completely mixed along the p-n boundary before they finally return to the voltage sensing reservoirs (Fig. 1(c)) where at contact pair 2-4 zero voltage drop is measured. At contact pair 6-8 the Landauer-Büttiker formalism predicts a Hall resistance of $R_H = h/e^2(1/|\nu_1| - \text{sgn}(\nu_1\nu_2)/|\nu_2|)$, with filling factors $\nu_1$ and $\nu_2$ in adjacent regions. Similarly the resistances in unipolar Hall-bars of type n-n$^+$ or p-p$^+$ are predicted by the ESM,[14,16] and the predictions are summarized in the scheme FIG. 1(e). Indeed, this scenario was now observed for current levels where pure edge state transport is not prevalent any more, and the results are reproduced in Fig. 3(a). The Hall plateau at $2\cdot h/2e^2$, resulting from regions with filling factors $\nu=\pm 2$, makes the p-n device useful for series resistance standards, whereas the plateaus at $h/3e^2$ (resulting from combining $|\nu|=2$ and $|\nu|=6$) and at $h/15e^2$ (resulting from combining $|\nu|=6$ and $|\nu|=10$) are probably less useful for application. All observed plateau sequences during a back gate voltage sweep may be related to the theoretical quantum resistance values derived within the ESM, and the ones we observed are indicated by the dashed line in scheme FIG. 1(e).

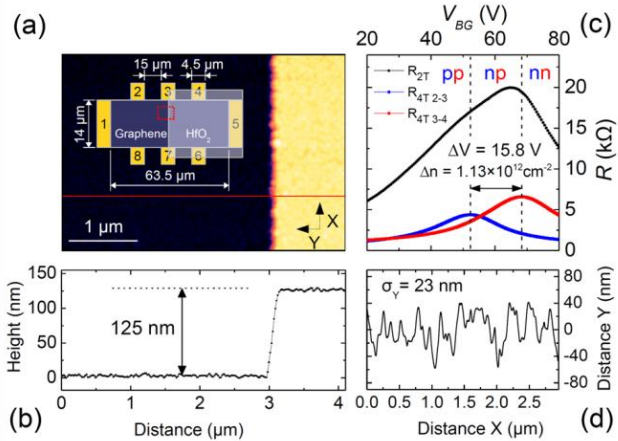

FIG. 2. (Color online) (a) AFM detail view and scheme (inset) of graphene Hall bar covered with HfO$_2$ on the right half. The Hall bar has eight contacts labelled 1-8. The dotted rectangle indicates where the AFM picture was taken. (b) AFM profile along the horizontal line in (a). (c) Two-terminal resistance between contacts 1 and 5 (upper curve) and four-terminal resistances along the uncovered and covered halves of the Hall-bar (lower curves) between contacts 2-3 (left curve) and contacts 3-4 (right curve) measured at 1 μA DC-current. (d) AFM profiles across the HfO$_2$ edge and on top of the HfO$_2$ layer taken along directions X,Y indicated in the lower right corner of (a).

The Landauer-Büttiker ESM assumes near-equilibrium linear transport,[17] a condition not met when current levels in the range of several microamperes are used, as required for resistance standards and as used in our experiment. However, the equivalent circuit model (ECM), applied to series quantum Hall devices, allows an alternative explanation of the obtained results, valid also in the high current regime. To apply the ECM, we considered the p-n junction FIG. 1(c) as a degenerate case of the device in FIG. 1(b) with a very high number $m$ of connections. The transition region between p and n-type regions is resistive, since it divides two different income-pressible Hall regions. The interconnects form bridges between these regions, much like the leads in Fig. 1(b). It is difficult to estimate theuirn resistance and hence $\varepsilon$, but even when the condition $\varepsilon \ll 1$ is relaxed to $\varepsilon < 1$, the voltage drop $V_{2-4}$, scaling with $(\varepsilon /(\varepsilon+2))^m$, would still vanish for large $m$. Since $V_{8-6} = V_{8-2} + V_{2-4} + V_{4-6}$ one obtains $V_{8-6} = 2 \cdot h/2e^2$ for filling factor ±2 in the p- and the n-region of the device. We verified that the ECM is not only applicable to the p-n junction case, but that it also predicts for the case of filling factors p-p$^+$ and n-n$^+$ the same Hall plateau sequence as the ESM. The ECM does even go beyond the ESM since for the case of equal carrier polarities it predicts a less rapid drop of the current density along the boundary line as for the p-n case. As we have no quantitative information about $\varepsilon$ and $m$, however, this prediction is only of qualitative character.

The best one can do is to resort to an experimental proof of the applicability of the concept, and to this end we performed a check of how well the condition $V_{2-4} = 0$ is met at increasing current levels. The results are presented in FIG 3(b) which shows the 4-terminal resistance between contacts 2 and 4 for current levels between 1 and 9 µA. Since the longitudinal zero-resistance plateaus were degraded by excess noise above gate voltages of 55 V, likely originating from a not perfect contact 2, we averaged $R_{2-4}$ only over the less noisy interval 50...55 V. We did indeed find that $<R_{2-4}> \approx 0$ within the uncertainty indicated by the error bars, see inset of FIG 3(b). The Hall resistances $R_{8-6}$ in the same current range are shown in FIG 3(c), and they agree with $2 \cdot h/2e^2$ within the scatter of the measured values over a finite gate voltage range up to 5 µA current. Note that in our device the coincidence on the gate voltage scale of the filling factor 2 plateaus for the p- and the n-type device regions was somewhat fortuitous since the plateau width at the chosen magnetic field nicely matched the interval between the CNPs of the two regions. This resulted from the different chemical doping in the covered and uncovered parts and was thus not influenced by us. Top-gating electrodes would be used in a carefully designed device to deliberately tune the Hall bar series array to the desired filling factors. Also note that the case of two p-n junctions in series can be readily extended to in principal any number of junctions according to the scheme illustrated in FIG. 1(f), where it is assumed that in alternating p- and n-regions a filling factor of 2 is established by appropriately controlled gates. Hall resistances of $M \cdot h/2e^2$, $M=2,3,4,5,...$ will be measured between the common side contact in the lower right corner and the contacts denoted by arrows. Contact pairs where the voltage drop is zero are also indicated. If the top gate voltages would not be controlled by one common but several different voltage sources, one could even switch the resistance standard between different values of $M$.

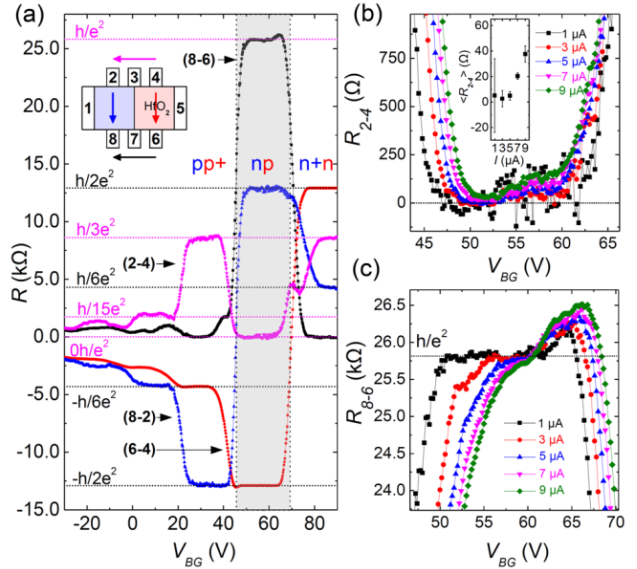

FIG 3. (color online) (a) Four-terminal resistance measurements in dependence on back-gate voltage obtained with 1 µA current between terminals 1 and 5. Labels at the curves denote at which contact pair they were measured. (b) Resistance measurements with currents from 1 to 9 µA measured at contact pair 6-8, and (c) at contact pair 4-2. The inset in (b) shows averaged resistance values and standard deviations in dependence on current. They were obtained by averaging over the back-gate voltage range where the flat minimum of the curves is not affected by excess noise.

In summary, we have shown that an array of graphene p-n junctions has the potential to be used as an interconnect-less quantum Hall resistance standard which realizes multiples of the resistance value $h/2e^2$. Predicted within the Landauer-Büttiker edge-state picture, this behaviour can also be derived when the equivalent circuit model of quantum Hall devices is extended to Hall bars with a quasi-continuum of interconnects. The applicability of the p-n junction series device concept for realistically high current levels was demonstrated.

This research has received funding from the European Community's Seventh Framework Programme, ERA-NET Plus, under Grant Agreement No. 217257.